\documentstyle[12pt,a4]{article}

\begin{document}

\title{BRST Field Theory of Relativistic Particles}

\author{J.W. van Holten\thanks{Research supported by the Stichting F.O.M.} \\
        NIKHEF-H, Amsterdam (NL)}

\newcommand{\nc}[2]{\newcommand{#1}{#2}}
\newcommand{\ncx}[3]{\newcommand{#1}[#2]{#3}}
\ncx{\pr}{1}{#1^{\prime}}
\nc{\nl}{\newline}
\nc{\np}{\newpage}
\nc{\nit}{\noindent}
\nc{\be}{\begin{equation}}
\nc{\ee}{\end{equation}}
\nc{\ba}{\begin{array}}
\nc{\ea}{\end{array}}
\nc{\dsp}{\displaystyle}
\nc{\bit}{\bibitem}
\nc{\ct}{\cite}
\ncx{\dd}{2}{\frac{\partial #1}{\partial #2}}
\nc{\pl}{\partial}
\nc{\dg}{\dagger}
\nc{\ag}{\alpha}
\nc{\bg}{\beta}
\nc{\gam}{\gamma}
\nc{\Gam}{\Gamma}
\nc{\bgm}{\bar{\gam}}
\nc{\del}{\delta}
\nc{\Del}{\Delta}
\nc{\eps}{\epsilon}
\nc{\ve}{\varepsilon}
\nc{\th}{\theta}
\nc{\vt}{\vartheta}
\nc{\kg}{\kappa}
\nc{\lb}{\lambda}
\nc{\Lb}{\Lambda}
\nc{\ps}{\psi}
\nc{\Ps}{\Psi}
\nc{\sg}{\sigma}
\nc{\spr}{\pr{\sg}}
\nc{\Sg}{\Sigma}
\nc{\rg}{\rho}
\nc{\fg}{\phi}
\nc{\Fg}{\Phi}
\nc{\vf}{\varphi}
\nc{\og}{\omega}
\nc{\Og}{\Omega}
\ncx{\nj}{1}{ \mbox{\boldmath $n_{#1}$} }
\nc{\Kq}{\mbox{$K(\vec{q},t|\pr{\vec{q}\,},\pr{t})$ }}
\nc{\Kp}{\mbox{$K(\vec{q},t|\pr{\vec{p}\,},\pr{t})$ }}
\nc{\vq}{\mbox{$\vec{q}$}}
\nc{\qp}{\mbox{$\pr{\vec{q}\,}$}}
\nc{\vp}{\mbox{$\vec{p}$}}
\nc{\va}{\mbox{$\vec{a}$}}
\nc{\vb}{\mbox{$\vec{b}$}}
\nc{\Ztwo}{\mbox{\sf Z}_{2}}
\nc{\Tr}{\mbox{Tr}}
\nc{\lh}{\left(}
\nc{\rh}{\right)}
\nc{\cB}{\mbox{$^{\ast}\Og$ }}
\nc{\nil}{\emptyset}
\nc{\bor}{\overline}
\renewcommand{\thepage}{}

\maketitle

\begin{abstract}
A generalization of BRST field theory is presented, based on wave operators for
the fields constructed out of, but different from the BRST operator. We discuss
their quantization, gauge fixing and the derivation of propagators. We show,
that the generalized theories are relevant to relativistic particle theories in
the Brink-Di Vecchia-Howe-Polyakov (BDHP) formulation, and argue that the same
phenomenon holds in string theories. In particular it is shown, that the naive
BRST formulation of the BDHP theory leads to trivial quantum field theories
with
vanishing correlation functions.
\end{abstract}

\np

\renewcommand{\thepage}{\arabic{page}}
\setcounter{page}{1}

\nit
1. {\bf Introduction}\nl

\nit
The dynamics of relativistic point particles in external electro-magnetic and
gravitational fields is characterised by the mass-shell condition

\be
g^{\mu\nu} \left( p_{\mu} - q A_{\mu} \right) \left( p_{\nu} - q A_{\nu}
 \right) + m^{2} c^{2} = 0,
\label{1}
\ee

\nit
where $A_{\mu}$ is the electro-magnetic vector potential, and $g^{\mu\nu}$
the inverse space-time metric. For particles with spin, eq.(\ref{1}) is
modified \ct{SkSt}-\ct{Krip}, but this modification is not essential for this
paper. Therefore we consider scalar particles only.

Although in classical point-particle mechanics eq.(\ref{1}) is a consequence of
the equations of motion, in field theory it is the starting point for defining
the dynamics of the corresponding fields: the classical field equations are
obtained by re-interpreting the quantities appearing in this equation as linear
operators. These operators act on fields taking values in the some
representation of the Lorentz group; for reasons mentioned above, in this paper
we discuss scalar fields only. Finally, the fields themselves may be taken to
represent the one-particle states of a corresponding relativistic quantum field
theory.

It is well-known, that equations like (\ref{1}) arise in classical mechanics as
first-class constraints in a theory with local reparametrization invariance
\ct{CGL}-\ct{BCL}. Therefore the corresponding operator equations can be
realized
in terms of a BRST operator $\Og$, which is nilpotent:

\be
\Og^{2} = 0,
\label{2.1}
\ee

\nit
and the cohomology classes of which correspond to the solutions of the
classical
field equations\footnote{See \ct{KO,MH} and references therein}:

\be
\left\{ \Og\, \Ps = 0\: \wedge\: \Ps \neq \Og\, \Lb \right\}
                      \Leftrightarrow {\cal H} \Ps = 0.
\label{3}
\ee

\nit
Here $\Ps$ is the classical field (or the corresponding one-particle quantum
state vector), and $\cal H$ is the appropriate wave operator --in this case the
Klein-Gordon operator, with proper inclusion of the coupling to
electro-magnetic
and gravitational background fields.

The definition of physical fields in terms of BRST cohomology, as in
eq.(\ref{3}), can be expressed naturally in terms of a classical variational
principle. Define an action

\be
S_{0} = 1/2\, \left( \Ps, \Og \Ps \right)
\label{4}
\ee

\nit
where $(\Fg,\Ps)$ denotes an inner product w.r.t.\ which $\Og$ is
self-adjoint\footnote{Since $\Og$ is nilpotent, this inner product necessarily
has an indefinite signature \ct{MS,JW3}.}; then $S_{0}$ is real and moreover
invariant under BRST transformations:

\be
\Ps \rightarrow \pr{\Ps} = \Ps\, +\, \Og \Lb,
\label{5}
\ee

\nit
with $\Lb$ an arbitrary field taking values in the same space as $\Ps$. The
invariance results from the nilpotency of the BRST operator, eq.(\ref{2.1}).
The
variation of the action (\ref{4}) vanishes precisely for those fields $\Ps$
which satisfy eqs.(\ref{3}), and which are defined only modulo a transformation
of type (\ref{5}).

A quantum field theory, the one-particle states of which satisfy these
BRST-type of field equations, can now be constructed by defining a generating
functional for Green's functions as

\be
Z\left[J\right] = \int \left[ D\Ps \right]\,
                       \exp i\left\{ S_{0} - (J,\Ps) \right\}.
\label{6}
\ee

\nit
Taking the integral naively as the integral over all classical configurations
$\Ps$ would of course lead to a divergent expression, due to the BRST gauge
invariance (\ref{5}). Hence an appropriate gauge fixing procedure modifying the
integration measure must be introduced to make the functional integral
well-defined. Note that physical sources must themselves be BRST-invariant:

\be
\Og\, J_{phys} = 0.
\label{6.1}
\ee

\nit
Finally, BRST-invariant self-interaction terms for the fields $\Ps$ can be
added
to the action (\ref{4}). Eq.(\ref{6}) thus provides a starting point for a
BRST-covariant quantum field theory. In particular, it can be used to set up
BRST-covariant perturbation theory.

In this paper we discuss the BRST construction of point-particle field theories
in quite some generality and show, that it contains several ambiguities which
need to be resolved before a well-defined theory is obtained. The procedure
described here was introduced originally in the construction of string field
theories\ct{WS,EW}, but its implementation in the field theory of point
particles has received less attention\footnote{See however, refs.\
\ct{WSB,HH}}.
The formal aspects of our results may be of some relevance in the context of
string field theory as well. \nl

\nit
{\bf 2. Gauge fixing} \nl

\nit
In this section we take the BRST-operator $\Og$, which specifies the field
content and dynamics of the physical system, as given and discuss the
implementation of the procedure to construct the generating functional for
Green's functions of the quantum field theory, as sketched in the introduction.

First we note that the action (\ref{4}) is not the most general BRST-invariant
quadratic action one can construct given the inner product $(\Fg,\Ps)$. Rather,
we can define a whole class of BRST-invariant actions specified by an
additional operator $G$ with the property

\be
G\, \Og = \left( G\, \Og \right)^{\dagger} = \Og\, G^{\dagger},
\label{7}
\ee

\nit
where a dagger denotes the adjoint w.r.t.\ the inner product introduced above.
The action

\be
S_{G} = 1/2\, \left( \Ps, G \Og \Ps \right)
\label{8}
\ee

\nit
is then both BRST-invariant and real. Of course, as long as $G$ is invertible,
we can redefine the fields so as to bring the action back into the form
(\ref{4}). However, $G$ may also be singular (for example, if it is a
projection
operator) and then the expression (\ref{8}) can not be rewritten in this form.
A situation of this kind is actually encountered in the examples discussed
below.

As mentioned, the BRST-invariance makes the naive functional integral

\be
Z_{G}\left[ J \right] = \int \left[ D\Ps \right]\, \exp i \left\{ S_{G}
                                                   - (J,\Ps) \right\}
\label{9}
\ee

\nit
ill-defined. In order to construct a better-defined quantum theory one must
restrict the integration to BRST-inequivalent configurations $\Ps$. A procedure
for finding such gauge-fixed configurations was described in refs.\
\ct{Ger}-\ct{HH}. Briefly, the idea is to introduce a nilpotent co-BRST
operator
$^{*}\Og$ with

\be
\ba{ll}
(i) & ^{*}\Og^{2} = 0, \\
  &  \\
(ii) & \left\{ \Og, \,^{*}\Og \right\} = \Del, \\
\ea
\label{10}
\ee

\nit
where we suppose the inverse $\Del^{-1}$ to exist in the space of fields
interacting with external sources. It also follows from the graded Jacobi
identity that

\be
\left[ \Del, \Og \right] = 0, \hspace{1cm} \left[ \Del, \,^{*}\Og \right] =0.
\label{11}
\ee

\nit
The definition of $^{*}\Og$ is not unique; in fact one can generally find
as many different co-BRST operators as there are different choices of gauge
for the underlying reparametrization symmetry \ct{JW3}, \ct{HH}.

Now we fix the BRST invariance by restricting the functional integral to field
configurations such that

\be
G\, ^{*}\Og\, \Ps = 0.
\label{13}
\ee

\nit
Reality of this condition requires the gauge-fixing operator to be hermitean:

\be
G\, ^{*}\Og =\,^{*}\Og\, G^{\dagger},
\label{13.1}
\ee

\nit
cf.\ eq.(\ref{7}). The invertibility of $\Del$ then implies that this gauge is
well-defined (it intersects each BRST orbit once). Namely, the classical field
equation obtained by varying the effective action in eq.(\ref{9}) is

\be
G \Og \Ps = J,
\label{14}
\ee

\nit
which requires $J$ to be BRST-invariant for consistency. Now the restriction
(\ref{13}) together with eq.(\ref{13.1}) implies

\be
^{*}\Og \Og G^{\dagger}\, \Ps = \Del G^{\dagger}\, \Ps =\, ^{*}\Og\, J,
\label{15}
\ee

\nit
and this equation has the solution

\be
G^{\dagger} \Ps = \Del^{-1}\, ^{*}\Og\, J.
\label{16}
\ee

\nit
{}From this result we directly obtain the classical (i.e., tree-level)
propagator
in the singular gauge, the analogue of the Landau gauge in electro-dynamics.
Again, we distinguish two cases: if $G$ has an inverse, we immediately have

\be
<\Ps \Ps>_{tree}\, =\, \frac{\del \Ps}{\del J}\,
                   =\, \frac{1}{\Del G^{\dagger}}\, ^{*}\Og.
\label{17}
\ee

\nit
On the other hand, if $G$ is not invertible, consistency requires that we can
write

\be
J = G K.
\label{18}
\ee

\nit
In this case we obtain a tree-level propagator for the field $\Fg =
G^{\dagger} \Ps$ as follows:

\be
<\Fg \Fg>_{tree}\, =\, \frac{\del \Fg}{\del K}\,
                   =\, \frac{1}{\Del}\, ^{*}\Og G.
\label{19}
\ee

\nit
In the functional integral the gauge condition can be implemented as
usual by inserting a delta-functional $\del[ G\, ^{*}\Og \Ps]$ in the
functional
integral, accompanied by the corresponding Faddeev-Popov determinant.
According to the standard Faddeev-Popov-'t Hooft prescription including a
Lagrange multiplier field $\Sg$, this determinant can be written using
Grassmann-odd ghost fields $(B,C)$ in the usual way, leading to an expression
of the type

\be
\ba{ll}
Z_{G}\left[ J \right] = & \dsp{ \int \left[ D\Ps \right] \left[ D\Sg \right]
     \left[ DB \right] \left[ DC \right]\, }\\
  &  \\
  &  \dsp{ \exp i \left\{ S_{G} - (J,\Ps) - \frac{\ag}{2}\, \left( \Sg, T G \Sg
     \right) - (\Sg, T G \,^{*}\Og \Ps) - (B, T G \,^{*}\Og \Og C) \right\}.
}\\
\ea
\label{20}
\ee

\nit
where $\ag T$ is a gauge-fixing operator to be chosen so as to make the
full effective quantum action well-defined. Shifting the field $\Sg$ to

\be
\Sg^{\prime} = \Sg + \frac{1}{\ag}\, G\, ^{*}\Og\, \Ps,
\label{20.1}
\ee

\nit
we can rewrite this in the form

\be
\ba{ll}
Z_{G}\left[ J \right] = & \dsp{ \int \left[ D\Ps \right] \left[ D\Sg^{\prime}
     \right] \left[ DB \right] \left[ DC \right]\, }\\
   &  \\
   &  \dsp{  \exp i \left\{ S_{G} + \frac{1}{2\ag}\, (\Ps, \,^{*}\Og T G
      \,^{*}\Og \Ps) - \frac{\ag}{2}\, \left( \Sg^{\prime}, T G \Sg^{\prime}
      \right) - (B, T G\, ^{*}\Og \Og C) - (J,\Ps) \right\}. }\\
\ea
\label{20.2}
\ee

\nit
Unfortunately, this expression is again ill-defined, because now the ghost
action is invariant under BRST-transformations of the ghosts $(B,C)$. Therefore
one also needs to restrict the integration over the ghost fields in the same
way, leading in the end to an infinite tower of ghost-for-ghosts with
alternating Grassmann parity. However, at least if there are no additional
self-interactions of the fields $\Ps$ we see that the functional integral over
$\Ps$ and the ghosts $(B,C)$ factorizes. Now define the operator $E$ by

\be
G E = \left\{ ^{*}\Og, T \right\} G;
\label{20.3}
\ee

\nit
in the examples in this paper one can actually choose the gauge-fixing operator
$T$ in such a way that $E = 1$. Provided $G$ is invertible we can now write the
integral over the $\Ps$ fields as

\be
Z_{G}\left[ J \right] = Z_{G}\left[ 0 \right]\, \exp -\frac{i}{2}\,
                        \left( J, M^{-1} G^{-1} J \right),
\label{21}
\ee

\nit
where

\be
M = \Og + \frac{1}{\ag}\, E\, ^{*}\Og,
\label{21.1}
\ee

\nit
and

\be
M^{-1} = \frac{1}{\Del}\,\left(\, ^{*}\Og + \Og \frac{\ag}{E} \right).
\label{21.2}
\ee

\nit
Note, that in the limit $\ag \rightarrow 0$ we reobtain the propagator
(\ref{17}), as anticipated. If $G$ is not invertible, we have to replace this
expression by

\be
Z_{G}\left[ K \right] = Z_{G}\left[ 0 \right]\, \exp -\frac{i}{2}\,
                        \left( K, G M^{-1} K \right).
\label{22}
\ee

\nit
If self-interactions of the $\Ps$ fields are added to the action, these
expressions can still be used as the starting point for BRST-invariant
perturbation theory in the standard fashion: replace the interactions
of the $\Ps$ by vertex operators, obtained by substitution of the functional
derivatives $\del /\del J$ for each $\Ps$ in the interaction term. In this
context we remark, that in general the introduction of self-interactions
modifies the BRST transformation rules so as to become $\Ps$-dependent. This
also introduces interactions of $\Ps$ with the ghost fields $(B,C)$, which have
to be treated similarly.
\nl\nl

\nit
{\bf 3. Point particles} \nl

\nit
The classical Lagrangian for a scalar point particle with mass $m$ and charge
$q$ moving in external fields $g_{\mu\nu}(x)$ and $A_{\mu}(x)$ is

\be
L = \frac{m}{2e}\, g_{\mu\nu} \dot{x}^{\mu} \dot{x}^{\nu} +
    q A_{\mu} \dot{x}^{\mu} - \frac{1}{2} m e.
\label{24}
\ee

\nit
Here $x^{\mu}$ are the position co-ordinates, a dot denotes the proper-time
derivative and $e$ is the einbein variable which makes the theory
reparametrization invariant on the world-line. We have also taken the velocity
of light to be unity. In principle we could eliminate $e$ as a redundant degree
of freedom in favor of a non-linear version of the theory, by replacing the
einbein by the expression

\be
e = \sqrt{ - g_{\mu\nu} \dot{x}^{\mu} \dot{x}^{\nu} },
\label{25}
\ee

\nit
as suggested by its classical equation of motion. However, in order to keep
the analogy with higher-dimensional reparametrization invariant theories --like
Einstein gravity-- in which the base-space metric has dynamical degrees of
freedom as close as possible, we prefer to keep the einbein in the theory as
an independent variable. In addition, in this formulation the Lagrangian
remains a quadratic function of the velocities.

In the formulation (\ref{24}), the theory has two first-class constraints
originating in the local world-line reparametrization invariance:

\be
{\cal H}_{0} = 0, \hspace{1.5cm} p_{e} = 0.
\label{26}
\ee

\nit
Here ${\cal H}_{0}$ is $1/e$ times the world-line Hamiltonian:

\be
2m {\cal H}_{0} = g^{\mu\nu}(p_{\nu} - q A_{\mu}) (p_{\nu} - q A_{\nu}) +
m^{2},
\label{27}
\ee

\nit
which is identical with the constraint (\ref{1}) upon taking $c = 1$, and
$p_{e}$ is the momentum conjugate to $e$.

In order to implement the BRST construction we introduce real Grassmann-odd
variables {\boldmath $g$} $= (b,c)$ with anti-real canonical momenta {\boldmath
$\pi$} $= (\pi_{b}, \pi_{c})$, having graded Poisson brackets

\be
\left\{ \mbox{\boldmath $g, \pi$} \right\} = i \mbox{{\bf 1}}.
\label{28}
\ee

\nit
Then we define the BRST-generator as

\be
\Og = c {\cal H}_{0} + \pi_{b} p_{e}.
\label{29}
\ee

\nit
The Poisson brackets of this quantity with the canonical co-ordinate and
momentum variables generate BRST transformations in classical phase space,
which
leave ${\cal H}_{0}$ and the action as defined from the Lagrangian (\ref{24})
invariant. Moreover, the graded Poisson bracket of $\Og$ with itself vanishes:

\be
\left\{ \Og, \Og \right\} = 0,
\label{30}
\ee

\nit
showing that these BRST transformations in classical phase space are nilpotent.

In field theory we replace the phase-space variables by operators with
(anti-)commutators equal to $i$ times the corresponding Poisson brackets. In
the
co-ordinate picture this results in the substitutions

\be
\ba{ll}
\dsp{ p_{\mu} = \frac{-i}{\sqrt[4]{-g}} \dd{}{x^{\mu}} \sqrt[4]{-g}, } &
\dsp{ p_{e} = -i \dd{}{e}, } \\
  &  \\
\dsp{ \pi_{b} = - \dd{}{b}, } & \dsp{ \pi_{c} = - \dd{}{c}, }\\
\ea
\label{31}
\ee

\nit
whilst $(x^{\mu},e,b,c)$ denote multiplication by corresponding real
(anti-)commuting c-numbers. Then we have

\be
{\cal H}_{0} = \frac{1}{2m}\, \left( - D^{2} + m^{2} \right),
\label{32}
\ee

\nit
where $D^{2}$ is the $U(1)$ and general-co-ordinate invariant laplacian on
scalar functions, and

\be
\Og = c {\cal H}_{0} + i \frac{\partial^{2}}{\partial b \partial e}.
\label{33}
\ee

\nit
As the space of fields we take the polynomials in $(b,c)$ with co-efficients in
the Cauchy completion of the set of square-integrable scalar functions of
$(x^{\mu},e)$. We define the scalar components by the expansion

\be
\Ps(b,c) = \ps - i\, b \ps_{b} + c \ps_{c} - i\, cb \ps_{cb}.
\label{33.1}
\ee

\nit
On this space we introduce an inner product

\be
(\Ps, \Fg) = i \int_{-\infty}^{\infty} de \int_{-\infty}^{\infty} d^{n}x
             \int db \int dc\: \sqrt{-g}\, \bar{\Ps} \Fg,
\label{34}
\ee

\nit
where $\bar{\Ps}$ is defined as

\be
\bar{\Ps} = \ps^{*} + i\, b \ps_{b}^{*} + c \ps_{c}^{*} - i\, cb \ps_{cb}^{*},
\label{35}
\ee

\nit
the star denoting ordinary complex conjugation. With respect to this inner
product both ${\cal H}_{0}$ and $\Og$ are self-adjoint. In the following we
call
a field $\Ps$ {\em real} if it has real {\em components} $\ps_{i}$.

Noting, that

\be
\Og \Ps = \dd{\ps_{b}}{e} + c \left( {\cal H}_{0} \ps - \dd{\ps_{cb}}{e}
\right)
          - i\, cb {\cal H}_{0} \ps_{b},
\label{35.01}
\ee

\nit
it is now straightforward to show, that the non-trivial BRST cohomology classes
consist of fields of the form

\be
\Ps(b,c) = -i\, b \ps_{b},
\label{35.1}
\ee

\nit
which satisfy

\be
\ba{ll}
\left(- D^{2} + m^{2} \right) \ps_{b} = 0, & \dsp{ \dd{\ps_{b}}{e} = 0.} \\
\ea
\label{36}
\ee

\nit
The first equation is the standard field equation for the wave function of a
single scalar particle in external fields, whilst the second equation expresses
the reparametrization invariance of the physical states of the theory: any
worldline reparametrization (a redefinition of proper time) can be absorbed in
a
redefinition of $e$; since physical states $\Ps$ do not depend on $e$, they are
{\em ipso facto} invariant under these reparametrizations.

Finally it follows from eq.(\ref{35.01}) that the BRST transformations of
the field components are:

\be
\ba{ll}
\dsp{ \del \ps = \dd{\xi_{b}}{e},} & \dsp{ \del \ps_{b} = 0, }\\
  &  \\
\dsp{ \del \ps_{c} = {\cal H}_{0} \xi - \dd{\xi_{cb}}{e}, } &
   \dsp{ \del \ps_{cb} = {\cal H}_{0} \xi_{b}.} \\
\ea
\label{36.1}
\ee

\nit
Hence the $\ps_{b}$-component of the field is the only one inert under BRST
transformations. \nl
\nl

\nit
{\bf 4. Quantum theory of the scalar field}\nl

\nit
Finally we turn to the construction of a quantum theory for the scalar field.
Without undue loss of generality we may take this field to be real (in the
sense
explained above). As a first step we look for operators $G$ with the property
(\ref{7}). They are of the form

\be
G = \nj{} + i\, c \nj{c} + i \nj{b} \dd{}{b} + i\, c \nj{cb} \dd{}{b},
\label{37}
\ee

\nit
where $(\nj{},\nj{b})$ are taken to commute with ${\cal H}_{0}$, and $(\nj{},
\nj{c})$ with $p_{e}$. The term with $\nj{cb}$ is immaterial, because it
vanishes upon multiplication with $\Og$. Hence we may choose $\nj{cb} = 0$ from
the start. Furthermore we observe, that $G$ is invertible if and only if
$\mbox{\boldmath $n$}^{-1}$ exists.

As in eq.(\ref{8}) we now obtain an action leading to the equations of motion
(\ref{36}) by taking

\be
S_{G} = 1/2\, \left( \Ps, G \Og \Ps \right),
\label{38}
\ee

\nit
with the inner product as defined in eq.(\ref{34}). In terms of components
this gives:

\be
S_{G} = \int_{-\infty}^{\infty} de \int_{-\infty}^{\infty} d^{n}x \:
        \sqrt{-g}\,\left\{ \ps \nj{} {\cal H}_{0} \ps_{b} + \ps_{cb} \nj{}
    \dd{\ps_{b}}{e} - \frac{1}{2}\, \ps_{b}  \left( \nj{b} {\cal H}_{0} -
    i \nj{c} \dd{}{e} \right) \ps_{b} \right\}.
\label{39}
\ee

\nit
We observe, that the component $\ps_{c}$ does not occur in this expression.
Indeed, such a component can always be gauged away by a BRST transformation
leaving the action invariant.

As a result, if we define a BRST-invariant functional integral of the type
eq.(\ref{9}) the integration over this component is not damped by exponential
factors, and hence will give rise to a divergent expression, as anticipated
above. The solution to this problem has already been discussed, see
eqs.(\ref{10}-\ref{20}). The procedure outlined in sect.\ 2 is suitable if we
can find an appropriate co-BRST operator. Such a co-BRST operator is for
example

\be
\ba{ll}
^{*}\Og = & - \pi_{c} \\
  &  \\
          & \dsp{ \mapsto \dd{}{c}. }\\
\ea
\label{40}
\ee

\nit
The corresponding BRST laplacian is then

\be
\Del =  {\cal H}_{0},
\label{41}
\ee

\nit
which up to a factor $1/2m$ is just the Klein-Gordon operator.

In order to complete the construction of the effective quantum action
(\ref{20.2}), we also specify an operator $T$ by

\be
T = c.
\label{42}
\ee

\nit
As a consequence,

\be
E = 1.
\label{43}
\ee

\nit
With this choice the gauge-fixed kinetic operator in the effective quantum
action (\ref{20.2}) becomes

\be
M = \Og + \frac{1}{\ag}\, ^{*}\Og = c{\cal H}_{0} + i \frac{\partial^{2}}{
    \partial e \partial b}\, + \frac{1}{\ag}\, \dd{}{c}.
\label{44}
\ee

\nit
It follows that the gauge-fixed form of the propagator is

\be
M^{-1} = \frac{2m}{\left( - D^{2} + m^{2} \right)}\, \left( \dd{}{c} +
         i \ag \frac{\partial^{2}}{\partial e \partial b} \right) + \ag c.
\label{44.1}
\ee

\nit
The gauge fixing terms in the action read in components

\be
\ba{l}
\dsp{ \frac{1}{2\ag}\, \left(\Ps, G\, ^{*}\Og \Ps \right) = } \\
  \\
 \hspace{1.5cm} \dsp{
     \frac{1}{2\ag}\, \int_{-\infty}^{\infty} de \int_{-\infty}^{\infty}
     d^{n}x\, \sqrt{-g} \left\{ i \ps \nj{c} \ps_{cb} + i \ps_{b} \nj{c}
\ps_{c}
     + \ps_{c} \nj{} \ps_{cb} + \ps_{cb} \nj{} \ps_{c} + \ps_{cb} \nj{b}
     \ps_{cb} \right\}. } \\
\ea
\label{45}
\ee

\nit
Adding this term to $S_{G}$, eq.(\ref{39}), gives the full expression for the
scalar-field dependent terms in the effective action.

{}From the component results (\ref{39},\ref{45}) it is easy to see that
different
choices of $G$ lead to physically quite different results. As a first example,
consider the case $G = 1$, or

\be
\ba{cc}
\nj{} = 1, & \nj{b} = \nj{c} = 0.\\
\ea
\label{46}
\ee

\nit
Then the effective quantum action is

\be
S_{eff} = \int_{-\infty}^{\infty} de \int_{-\infty}^{\infty} d^{n}x \,
\sqrt{-g}
          \left( \ps {\cal H}_{0} \ps_{b} + \ps_{cb} \left( \dd{\ps_{b}}{e} +
          \frac{1}{\ag}\, \ps_{c} \right) + ghost\: terms \right).
\label{47}
\ee

\nit
It follows, that for $\ag \neq 0$ the functional integral for fields coupled to
physical (BRST-invariant) sources becomes, after integrating out
$(\ps, \ps_{c}, \ps_{cb})$:

\be
Z\left[ j \right] = {\cal N}\, \int [D\ps_{b}]\, \del \left[{\cal H}_{0}
                    \ps_{b} \right]\, \exp -i \int_{-\infty}^{\infty} de
                    \int_{-\infty}^{\infty} d^{n}x \,\sqrt{-g}\, j \ps_{b},
\label{48}
\ee

\nit
where $\cal N$ is a $\ps$-independent (but background field dependent)
normalization factor coming from the integral over the ghosts. Therefore only
classical configurations satisfying the Klein-Gordon equation contribute to the
functional integral, and quantum fluctuations are suppressed. In particular, in
this theory the 2-point function vanishes. One way to see this is to observe,
that writing the $\del$-functional as a Fourier integral the source term can be
absorbed by a shift of the multiplier field. Hence the right-hand side is
really
independent of the source $j$. Clearly, this theory is not equivalent with the
standard quantum theory of the scalar field.

In contrast, the standard scalar field theory is obtained for the singular
choice

\be
\ba{ll}
G & \dsp{ = c \dd{}{e} + 2im \dd{}{b} }\\
  &  \\
  & = - G^{\dagger},
\ea
\label{49}
\ee

\nit
which implies

\be
\nj{} = 0, \hspace{1cm} \nj{b} = 2m, \hspace{1cm} \nj{c} = p_{e} = - i\dd{}{e}.
\label{49.1}
\ee

\nit
With this choice one finds for the effective quantum action

\be
\ba{ll}
S_{eff} = & \dsp{ \int_{-\infty}^{\infty} de \int_{-\infty}^{\infty} d^{n}x \,
        \sqrt{-g}\, \left( - \frac{1}{2}\, \ps_{b} (2m {\cal H}_{0} -
        \dd{^{2}}{e^{2}}) \ps_{b} - \frac{1}{2\ag}\, \ps_{c} \dd{\ps_{b}}{e}\,
        \right. } \\
  &  \\
  &  \dsp{ \left. - \frac{1}{16\ag m}\, (\dd{\ps}{e})^{2} + \frac{m}{\ag}\,
     (\ps_{cb} - \frac{1}{4m}\, \dd{\ps}{e})^{2} + ghost \: terms \right). } \\
\ea
\label{50}
\ee

\nit
Integration over the $\ps_{cb}$, $\ps$ and $\ps_{c}$ now gives

\be
\ba{ll}
Z\left[ j \right] & \dsp{ = {\cal N}^{\prime}\, \int [D\ps_{b}]\, \del \left[
     \dd{\ps_{b}}{e}\right] \exp i \bg \int_{-\infty}^{\infty} de
     \int_{-\infty}^{\infty} d^{n}x \, \sqrt{-g}\, \left\{ - \frac{1}{2}\,
     \ps_{b} ( - D^{2} + m^{2} )\ps_{b}^{2} - j \ps_{b} \right\} } \\
  &  \\
  &  \dsp{ = \frac{{\cal N}^{\prime}}{\sqrt{ \det \left( -D^{2} + m^{2}
     \right)}}\, \exp \frac{i}{2} \int_{-\infty}^{\infty} d^{n}x\, \sqrt{-g}\,
     j \Del^{-1}_{KG} j, }\\
\ea
\label{51}
\ee

\nit
where $\Del^{-1}_{KG}$ denotes the Feynman propagator for the Klein-Gordon
scalar field, and a factor $\bg$ has been introduced to normalize the
integration\footnote{Effectively this represents a renormalization of $\hbar$.}
over $e$. In view of the $\del$-function contraint on $\ps_{b}$, it is
legitimate to take the source to be $e$-independent from the start. We observe
that although this theory does give the correct Green's functions for the
scalar
field, full equivalence with the standard scalar field theory is achieved only
if the normalization factor ${\cal N}^{\prime}$, which results from integration
over the ghosts, is independent of the background fields $(A_{\mu},
g_{\mu\nu})$.
This poses a strict constraint on possible ways to deal with the infinite tower
of ghost-for-ghosts in the present formulation of the theory.

Note that both choices of $G$ discussed above yield the correct classical
equations of motion. Nevertheless they correspond to different quantum
theories, because they lead to different Green's functions for the fields.
In order to investigate this point in some more detail, consider the class
of interpolating gauges

\be
G = \lb + \mu c \dd{}{e} + 2 i m \nu \dd{}{b},
\label{52}
\ee

\nit
where $(\lb,\mu,\nu)$ are numerical parameters. For $\lb = 1$ and $\mu = \nu
=0$
we reobtain the first case $G = 1$, whilst for $\lb = 0$ and $\mu = \nu =1$ we
get back the standard theory defined by the $G$ of eq.(\ref{49}). Inserting
this
into eqs.(\ref{39},\ref{45}) we first note, that for $\nu = 0$ and $\lb \neq 0$
we are always restricted to terms linear in the unphysical field components
$(\ps, \ps_{c}, \ps_{cb})$, which act as Lagrange multipliers in the functional
integral and always lead to a $\del$-function constraint requiring ${\cal
H}_{0}
\ps_{b} = 0$ in the measure of the functional integral, like in eq.(\ref{48}).
Therefore we assume $\nu \neq 0$.

Now we know already that in this case $\lb = 0$ leads to the correct standard
functional integral, eqs.(\ref{50},\ref{51}). Therefore we also take $\lb
\neq 0$. Then after performing the shift of fields

\be
\ba{lll}
\ps_{cb}^{\prime} & = & \dsp{ \ps_{cb}\, + \frac{\lb}{2m\nu}\, \ps_{c}\, -
   \frac{\mu}{4m\nu}\, \dd{\ps}{e}\, + \frac{\ag \lb}{2m\nu}
\dd{\ps_{b}}{e},}\\
  & & \\
\ps_{c}^{\prime} & = & \dsp{ \ps_{c}\, - \frac{\mu}{2\lb}\, \dd{\ps}{e}\, +
   \left( \ag + \frac{m\nu\mu}{\lb^{2}} \right) \dd{\ps_{b}}{e}, }\\
\ea
\label{53}
\ee

\nit
the effective action reads in components:

\be
\ba{ll}
\dsp{ S_{eff} = \int_{-\infty}^{\infty} de \int_{-\infty}^{\infty} d^{n}x
   \sqrt{-g}\, } & \dsp{ \left( \frac{m\nu}{\ag}\, \ps_{cb}^{\prime \, 2}\, -
   \frac{\lb^{2}}{4\ag m\nu}\, \ps_{c}^{\prime \, 2}\, - \frac{\nu}{2}\,
\ps_{b}
   {\cal H}_{eff} \ps_{b} \, \right. }\\
  &  \\
  & \dsp{ \left. + \frac{\lb}{2m}\, \ps {\cal H}_{eff} \ps_{b}
          +\, ghost\, terms \right). }\\
\ea
\label{54}
\ee

\nit
Here the effective wave operator is given by

\be
{\cal H}_{eff} = - D^{2} + m^{2} + \frac{m \mu^{2} }{2 \ag \lb^{2}}\,
                 \dd{^{2}}{e^{2}}.
\label{55}
\ee

\nit
The integration over the auxiliary field components $(\ps_{c}^{\prime},
\ps_{cb}^{\prime})$ is trivial, but the intergral over $\ps$ is now seen to
lead
{\em always} to a $\del$-function constraint ${\cal H}_{eff} \ps_{b} = 0$ in
the
functional integral. We conclude, that with the exception of the singular case
$\lb = 0$, $\nu \neq 0$ we always get trivial quantum field theories.
\nl

\nit
{\bf 5. Self-Interactions}\nl

\nit
The above actions can without much difficulty be extended to include
self-interactions of the scalar fields $\Ps$ \ct{HH}. The basic observation is,
that since the $\ps_{b}$ component is inert under BRST transformations, any
interaction involving only this component is automatically BRST-invariant
under the {\em original} set of BRST transformations. Note that this situation
differs from the case of string theories, in which both the action and the BRST
transformation rules have to be modified.

Restricting ourselves to the case of polynomial interactions, we find that
there are many different ways to write such interactions in terms of the
BRST superfields (\ref{33.1}). For the case of $G$ as in eq.(\ref{49}),
which is the only one of physical interest, a convenient choice is

\be
\ba{ll}
\Del S_{int} & \dsp{ = \frac{-g_{n}}{n!}\, \int_{-\infty}^{\infty} de
               \int_{-\infty}^{\infty} d^{n}x \int db \int dc\:
               cb\, \sqrt{-g}\, \left( \frac{G \Ps}{2m} \right)^{n} }\\
  &  \\
  &  \dsp{ = \frac{- g_{n}}{n!}\, \int_{-\infty}^{\infty} de
             \int_{-\infty}^{\infty} d^{n}x\: \sqrt{-g}\, \ps_{b}^{n}. }\\
\ea
\label{56}
\ee

\nit
Note, that in general the interacting theory does not take the form of
a Cherns-Simons like theory, as is the case of the open string. In particular,
in the action (\ref{50}) there are no {\em gauge independent} quadratic
auxiliary field terms which could be used to cast the polynomial interactions
for, say, the case $n = 4$ into cubic form.

Finally we remark that since in the interacting scalar theory the BRST
transformation rules are not modified, neither are the gauge-fixing and ghost
terms in the effective quantum action (\ref{50}).\nl

\nit
{\bf 6. The ghost terms}\nl

\nit
For the particular case of the theory defined by $G$ of eq.(\ref{49}), and with
the choice (\ref{40}) for $^{*}\Og$ and (\ref{42}) for $T$, the ghost terms in
the effective action become particularly simple and can be seen to result in a
background-independent factor in front of the functional integral, which may
subsequently be ignored.

Indeed, substituting the explicit expressions into eq.(\ref{20.2}) and
denoting the components of $(\Sg,B,C)$ by $(\sg_{i},\bg_{i},\gam_{i})$, one
obtains

\be
- \frac{\ag}{2}\, \left( \Sg^{\prime}, T G \Sg^{\prime} \right) -
   (B, T G\, ^{*}\Og \Og C) = \int_{-\infty}^{\infty} de
\int_{-\infty}^{\infty}
   d^{n}x\: \sqrt{-g}\, \left( -\ag m\, \sg_{b}^{2} + 2m\, \bg_{b} \gam_{b}
   \right).
\label{57}
\ee

\nit
Even though the functional integral over the other ghost-field components is
not damped, the result of further gauge fixing and the introduction of
ghost-for-ghosts will only result in an over-all factor which is
background-field independent and cancels from any physical Green's function in
the theory. In fact, this factor is BRST invariant by itself and may be divided
out from the start without spoiling BRST invariance of the theory.
\np

\nit
{\bf 7. Discussion}\nl

\nit
As illustrated by the example of the scalar point particle, the BRST
formulation
of a quantum field theory is not always obtained from the naive action $( \Ps,
\Og \Ps) + interactions $, but rather one needs the more general form
(\ref{8}).
In the case of the scalar particle the effective kinetic operator is

\be
\Og_{G} \equiv G \Og =
        i c\, \left( D^{2} + \dd{^{2}}{e^{2}} - m^{2} \right)\, \dd{}{b}.
\label{58}
\ee

\nit
The operator $\Og_{G}$ is a {\em Grassmann-even} nilpotent operator. Obviously,
transformations of the form $\del \Ps = \Og_{G} \Lambda$ leave the action
invariant and the physical states belong to the cohomology classes of
$\Og_{G}$.
However, the set of transformations generated by $\Og_{G}$ is smaller than the
BRST transformations generated by $\Og$ itself, which we already know to be a
full symmetry of the action $S_{G}$. Hence the cohomology of $\Og_{G}$ cannot
by itself completely determine the physical states. Indeed, the states obtained
from the cohomology of $\Og_{G}$ are zero modes of the effective wave operator

\be
{\cal H}_{eff} = - D^{2} - \dd{^{2}}{e^{2}} + m^{2} ,
\label{59}
\ee

\nit
but they are not necessarily annihilated by the einbein momentum operator
$p_{e} = -i\partial{}/\partial{e}$. Hence this really describes a particle in a
5-dimensional space-time, with $e$ acting as the fifth co-ordinate.

We may contrast this situation with the case in which we start from the
Nambu-Goto type theory \ct{CGL} in which the einbein $e$ is not an independent
variable, but rather given by the non-linear expression (\ref{25}). In that
case, there is only one independent first-class constraint, the mass-shell
condition $2m {\cal H}_{0} = (p - q A)^{2} + m^{2} = 0$. The corresponding BRST
operator is now the `irreducible part' of $\Og$:

\be
\Og_{irr} = c {\cal H}_{0}.
\label{60}
\ee

\nit
Since there is no gauge variable $e$ present, which has to be eliminated
from the physical states, we donot need to introduce more ghosts and we
find that the physical states are defined directly by the cohomology of
$\Og_{irr}$ in the space of polynomials $\Fg [c]$ in the ghost $c$ only.
In this space the BRST-invariant action for the corresponding quantum field
theory is simply

\be
\ba{ll}
S & = m \, \int_{-\infty}^{\infty} d^{n}x \int dc\, \sqrt{-g}\,
      \Fg \Og_{irr} \Fg \\
  &  \\
  & \dsp{ = 1/2 \int_{-\infty}^{\infty} d^{n}x\, \sqrt{-g}\, \fg \left(
          - D^{2} + m^{2} \right)\, \fg, } \\
\ea
\label{61}
\ee

\nit
where $\fg$ is the lowest component of $\Fg$, which is invariant under the
irreducible BRST transformations generated by $\Og_{irr}$.

Actually the same result is known to hold in the case of the open bosonic
string: the action of the type (\ref{61}) is known to give the correct
string field equations provided the BRST operator $\Og$ is constructed
from the dynamical Virasoro constraints only, as follows naturally starting
from the Nambu-Goto action. Starting from the Brink-Di Vecchia-Howe-Polyakov
action \ct{BDH}-\ct{AMP} with the 2-dimensional metric as independent gauge
degrees of freedom, one gets additional first-class constraints eliminating
these metric degrees of freedom from physical states. Then the BRST operator
contains extra terms constructed with the additional ghosts, and the action
principle needs modifications of the type we have discussed
here. \nl\nl

\end{document}